\documentclass[prl,twocolumn,showpacs,amsmath,amssymb]{revtex4}
\usepackage{bm}
\usepackage{graphicx}

\newcommand{\ud}{\mathrm{d}}
\newcommand{\ui}{\mathrm{i}}
\newcommand{\rX}{r_{\mathrm{X}}}

\newcommand{\myfigsep}{0.02 \textwidth}
\newcommand{\myfigscl}{0.44}
\newcommand{\myfigwid}{0.42 \textwidth}

\begin{document}

\title{Coherent Development of Neutrino Flavor
in the Supernova Environment}
\newcommand*{\UCSD}{Department of Physics, %
University of California, San Diego, %
La Jolla, CA 92093-0319}
\affiliation{\UCSD}
\newcommand*{\LANL}{Theoretical Division, Los Alamos National Laboratory, %
Los Alamos, NM 87545}
\affiliation{\LANL}
\newcommand*{\UMN}{School of Physics and Astronomy, %
University of Minnesota, Minneapolis, MN 55455}
\affiliation{\UMN}

\author{Huaiyu Duan}
\affiliation{\UCSD}
\author{George M.~Fuller}
\affiliation{\UCSD}
\author{J.~Carlson}
\affiliation{\LANL}
\author{Yong-Zhong Qian}
\affiliation{\UMN}

\date{\today}

\begin{abstract}
We calculate coherent neutrino and antineutrino flavor transformation
in the supernova environment, for the first time
including self-consistent 
coupling of intersecting neutrino/antineutrino trajectories.
For neutrino mass-squared difference 
$|\delta m^2|=3\times 10^{-3}\,\mathrm{eV}^2$
we find that in the normal (inverted) mass hierarchy
the more tangentially-propagating (radially-propagating)
neutrinos
and antineutrinos can initiate collective, simultaneous
medium-enhanced flavor conversion
of these particles across broad ranges of energy and
propagation direction.   
Accompanying alterations
in neutrino/antineutrino energy spectra and fluxes could affect
supernova nucleosynthesis and
the expected neutrino signal.
\end{abstract}
 

\maketitle

In this letter we present the first self-consistent solution to a long standing
problem in following coherent flavor inter-conversion among
neutrinos and antineutrinos in the region above 
the hot proto-neutron star subsequent to the supernova
explosion \cite{Duan:2006an}. The problem is that
flavor histories on intersecting neutrino/antineutrino world lines
can be coupled by neutral-current forward exchange scattering
\cite{Qian:1994wh}. Many studies neglecting this aspect of flavor
transformation 
\cite{Fuller:1987aa,Fuller:1992aa,Mezzacappa:1999co,%
Qian:1993dg,Qian:1994wh,Pastor:2002we,Qian:1995ua,%
Balantekin:2004ug,Fuller:2005ae}
nevertheless have shown that flavor conversion in the neutrino and
antineutrino fields above the proto-neutron star could
be important in understanding the supernova explosion mechanism
\cite{Fuller:1987aa,Fuller:1992aa,Mezzacappa:1999co}
and the origin of heavy \textit{r}-process nuclei
\cite{Qian:1993dg,Qian:1994wh,Pastor:2002we,Qian:1995ua,%
Balantekin:2004ug,Fuller:2005ae}.

\begin{figure}
\includegraphics*[width=0.4 \textwidth, keepaspectratio]{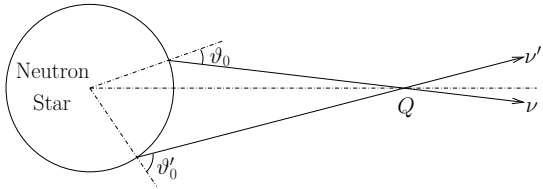} 
\caption{\label{fig:nubulb}
Flavor evolution of neutrino $\nu$ on trajectory designated by angle 
$\vartheta_0$
relative to the neutron star surface normal is coupled with the flavor
development of all neutrinos $\nu^\prime$ on intersecting trajectories.}
\end{figure}

\begin{figure*}
\begin{center}
$\begin{array}{@{}c@{\hspace{\myfigsep}}c@{}}
\includegraphics*[width=\myfigwid, keepaspectratio]{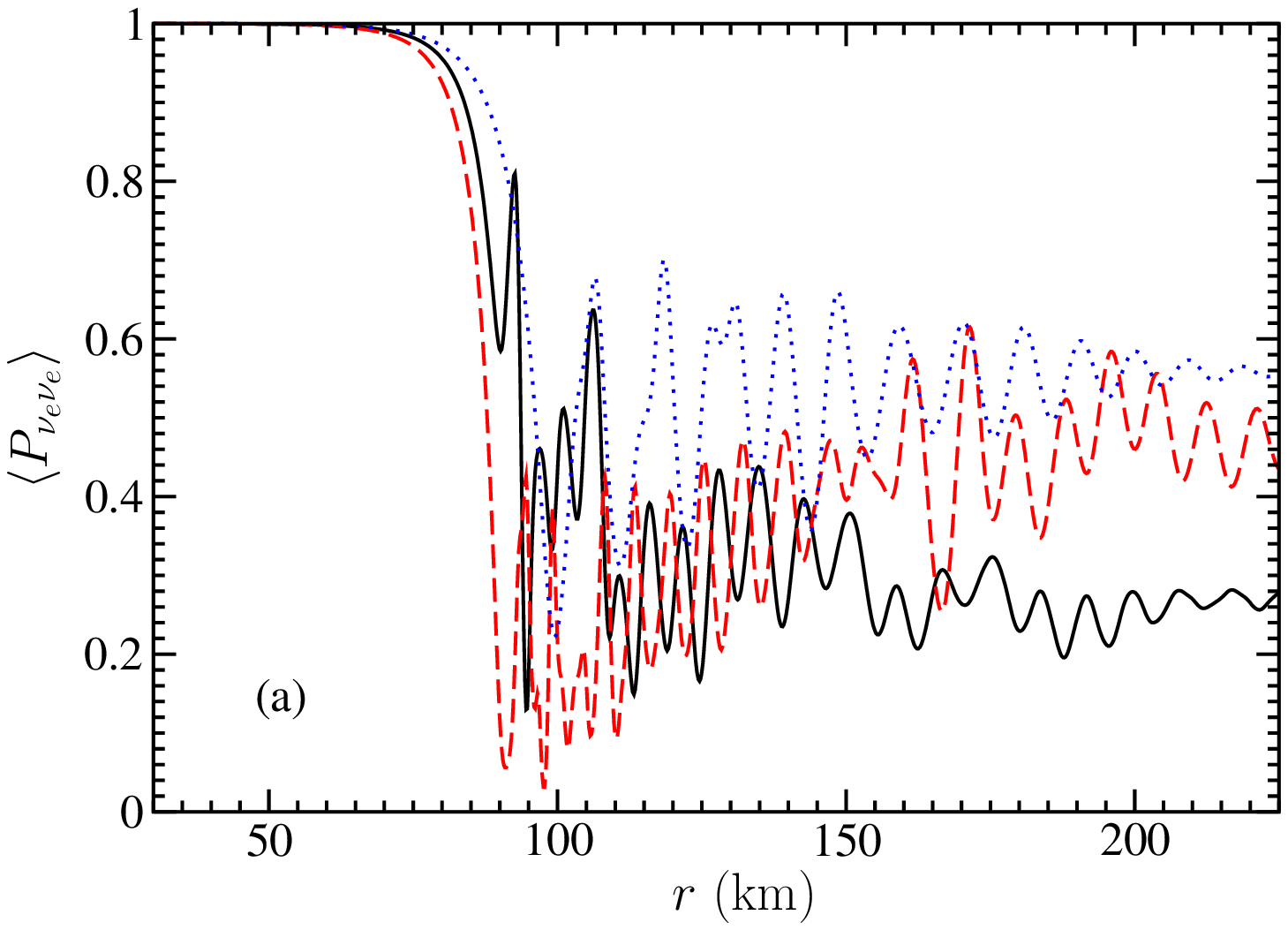} &
\includegraphics*[width=\myfigwid, keepaspectratio]{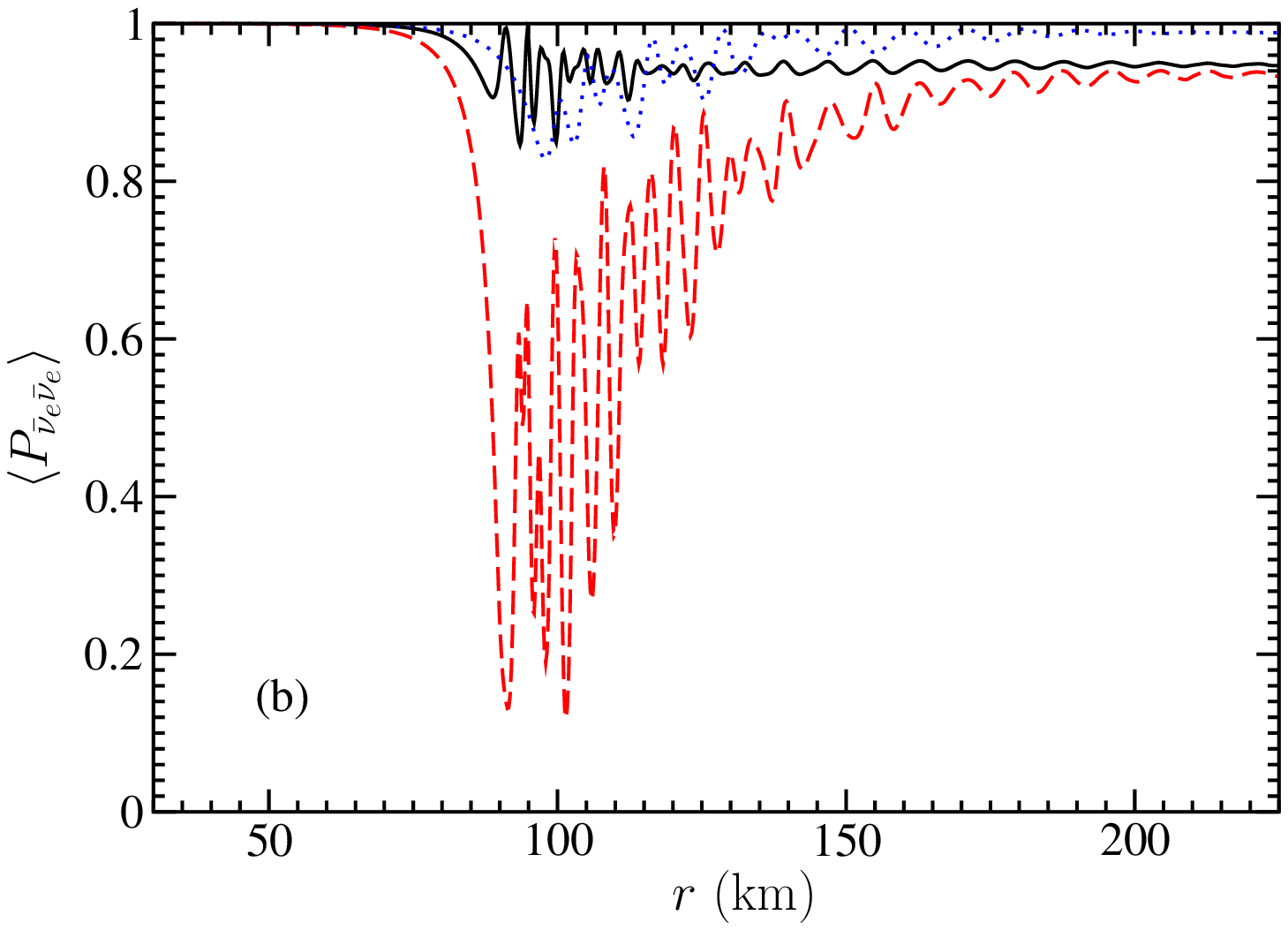} \\
\includegraphics*[width=\myfigwid, keepaspectratio]{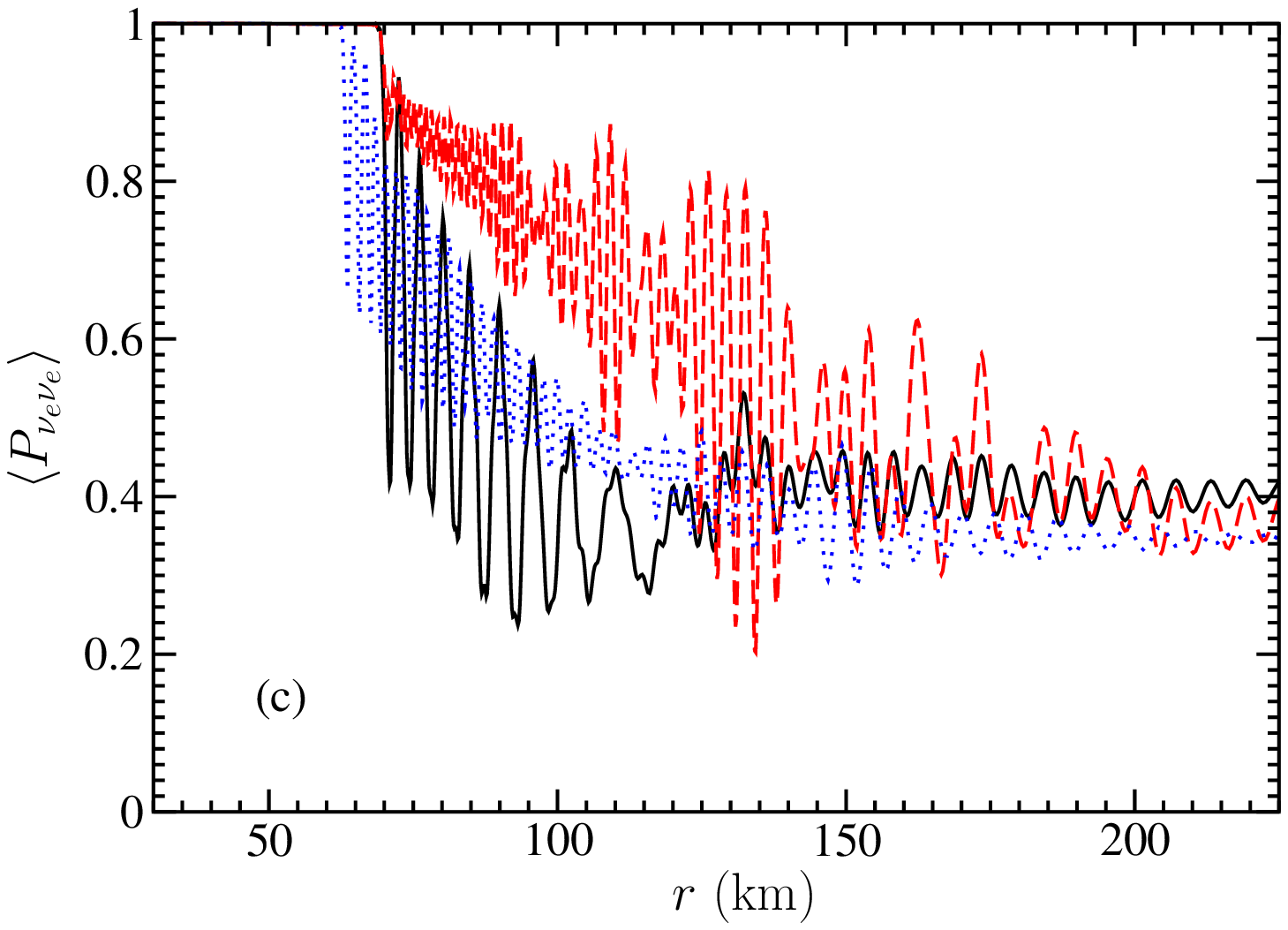} &
\includegraphics*[width=\myfigwid, keepaspectratio]{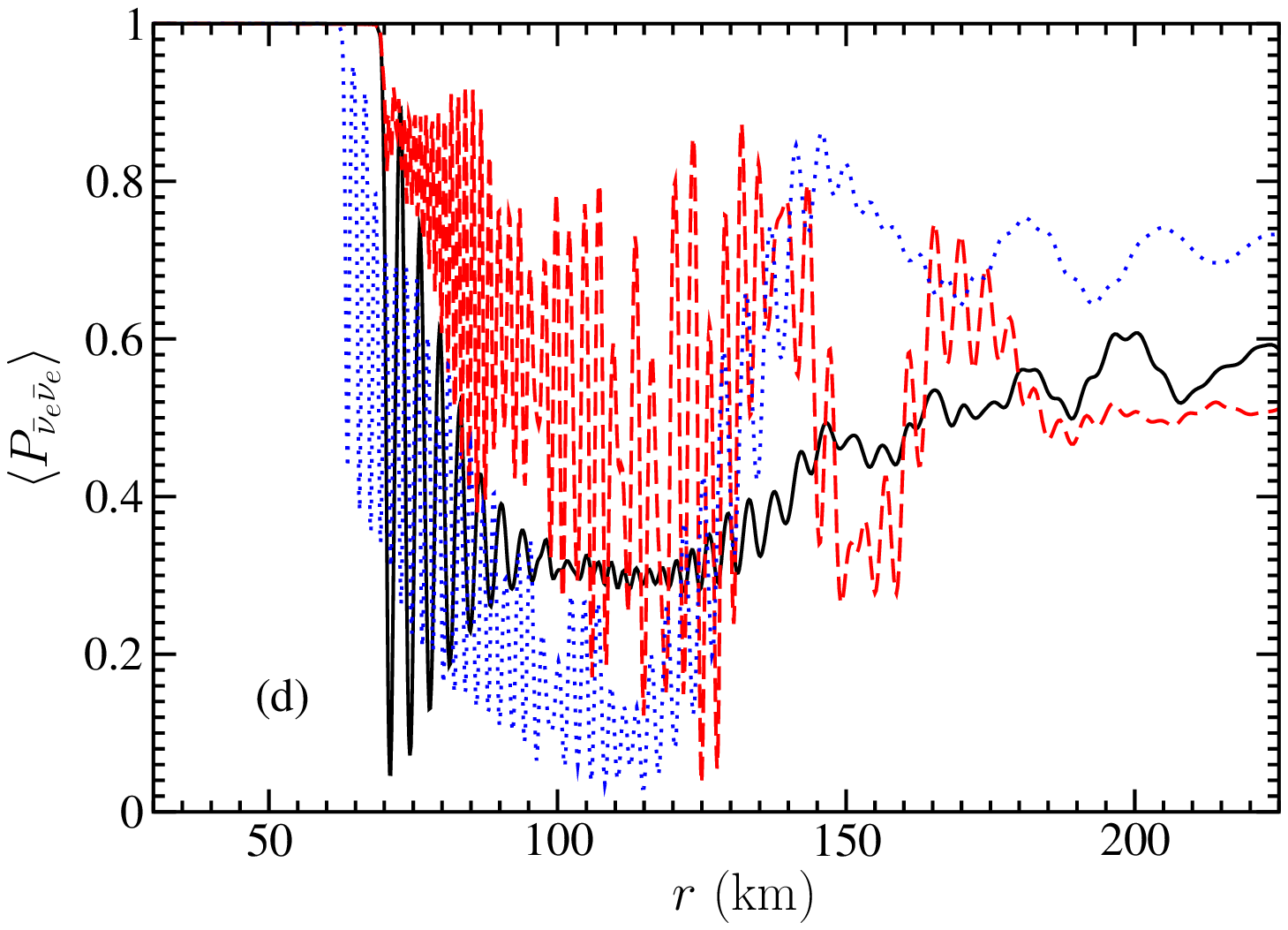}
\end{array}$
\end{center}
\caption{\label{fig:P-r}(Color online) 
Plots of energy-averaged survival probabilities $\langle P_{\nu\nu}\rangle$ 
for $\nu_e$ (left panels) 
and $\bar\nu_e$ (right panels) as functions of radius $r$
for the normal (upper panels) and
inverted (lower panels) neutrino mass hierarchies, respectively.
The solid and dashed lines give average survival probabilities
along the trajectories with $\cos\vartheta_0=1$ and $\cos\vartheta_0=0$,
respectively, as computed in the multi-angle simulations. The
dotted lines give the average survival probabilities
computed in the single-angle simulations.}
\end{figure*}

Inelastic scattering processes and associated de-coherence 
may dominate neutrino flavor
development in the proto-neutron star core and in the
region near the neutrino sphere, necessitating a full
quantum kinetic approach there \cite{Sigl:1992fn,Strack:2005ux}.
By contrast, in the hot bubble 
(a high-entropy region that develops above the neutrino sphere at
time post-core-bounce $t_{\mathrm{PB}}\gtrsim 3$ s), where 
the \textit{r}-process
elements may be made, neutrinos and antineutrinos for the most
part propagate coherently. In this limit we can model the evolution
of flavor along a {\it single} neutrino's trajectory with a mean field
\cite{Friedland:2003eh}, Schr\"{o}dinger-like equation, which 
in the effective $2\times 2$ mixing channel is
\begin{equation}
\label{eq:schroedinger-eq}
\ui \frac{\ud}{\ud t} \begin{pmatrix}
a_{e \alpha} \\
a_{\tau \alpha}
\end{pmatrix} 
=  {\hat{H}}_{\rm f}
 \begin{pmatrix}
a_{e \alpha} \\
a_{\tau \alpha}
\end{pmatrix}.
\end{equation}
The effective neutrino flavor
evolution Hamiltonian can be expressed in the 
flavor basis as
 \begin{equation}
 \label{eq:Ham-eq}
 {\hat{H}}_{\rm f} =
 \begin{pmatrix}
-\Delta\cos 2\theta + A + B & \Delta\sin 2\theta + B_{e\tau} \\
\Delta\sin 2\theta + B_{e\tau}^* & \Delta\cos 2\theta - A - B
\end{pmatrix}.
 \end{equation}
The flavor evolution of an antineutrino is determined similarly but
with $A\rightarrow -A$, $B\rightarrow -B$ and 
$B_{e\tau}\rightarrow -B_{e\tau}^*$. 
In these expressions $t$ is an Affine parameter along the neutrino's world
line, $\Delta\equiv \delta m^2/2E_\nu$, where $E_\nu$ is neutrino energy,
and $\delta m^2=m_3^2-m_1^2$ 
is the difference of the squares of the relevant vacuum neutrino
mass eigenvalues. 
We focus on the atmospheric mass-squared difference $\delta m_\mathrm{atm}^2$ 
because it will give flavor transformation deeper in the supernova
envelope than will the solar scale.
Therefore,
we set $\delta m^2=\pm 3\times{10}^{-3}\,{\rm eV}^2$, where the
plus (minus) sign is for the normal (inverted) mass hierarchy.
In Eq.\ \eqref{eq:Ham-eq}, $\theta$
is the effective $2\times 2$ vacuum mixing angle.
Flavor transformation in the $\nu_e\rightleftharpoons\nu_{\mu,\tau}$
and $\bar\nu_e\rightleftharpoons\bar\nu_{\mu,\tau}$ channels 
is most important in supernovae because there may be
disparities in energy spectra and fluxes among the neutrino flavors,
and because $\nu_e$ and $\bar\nu_e$ play a prominent role in
setting composition and in the prospects for signal detection.
In these channels $\theta\sim\theta_{13}$ for $\delta m_\mathrm{atm}^2$. 
Experiment suggests
 $\sin^22\theta_{1 3} \lesssim 0.1$ \cite{Fogli:2005cq}, and in
our numerical calculations we take $\theta = 0.1$.
The flavor diagonal potentials
in Eq.~\eqref{eq:Ham-eq} are $A$, from $\nu_e$-electron
forward scattering (determined by the matter density profile in the
hot bubble \cite{Duan:2006an}), and
$B$, from neutrino-neutrino forward scattering. The flavor off-diagonal
potential $B_{e\tau}$ similarly stems from neutrino-neutrino
forward exchange scattering \cite{Pantaleone:1992xh}.
Eq.~\eqref{eq:schroedinger-eq} is nonlinear in 
that the potentials $B$ and $B_{e\tau}$ 
depend on the amplitudes $a_{e \alpha}$ 
and $a_{\tau \alpha}$ for a neutrino with initial flavor state 
$\alpha = e,\tau$ to be either electron
or tau flavor, respectively. 
However, the true complexity of this problem arises from
quantum mechanical and geometrical coupling of neutrino/antineutrino
flavor histories as illustrated in Fig.~\ref{fig:nubulb}:
$B$ and $B_{e\tau}$ help determine the flavor development at point $Q$
on neutrino $\nu$'s world line, but these potentials depend on a 
coherent sum over all neutrinos $\nu^\prime$
passing through $Q$.
Here $\nu_\tau$ designates the relevant 
combination of mu and tau flavor
neutrinos assuming these species are maximally mixed in 
vacuum and in the supernova medium \cite{Balantekin:1999dx}. 
In our example numerical calculations we have taken
the initial neutrino and antineutrino
energy spectra to be of Fermi-Dirac
form, with degeneracy parameter $\eta_\nu=3$ and average energies
$\langle E_{\nu_e}\rangle = 11\,{\rm MeV}$,
$\langle E_{\bar\nu_e}\rangle = 16\,{\rm MeV}$,
and $\langle E_{\nu_\tau,\bar\nu_\tau}\rangle = 25\,{\rm MeV}$
and we take the energy luminosity for each
neutrino species to be $L_\nu = {10}^{51}\,{\rm erg}\,{\rm s}^{-1}$.  

\begin{figure*}
\begin{center}
$\begin{array}{@{}c@{\hspace{\myfigsep}}l@{}}
\includegraphics*[scale=\myfigscl, keepaspectratio]{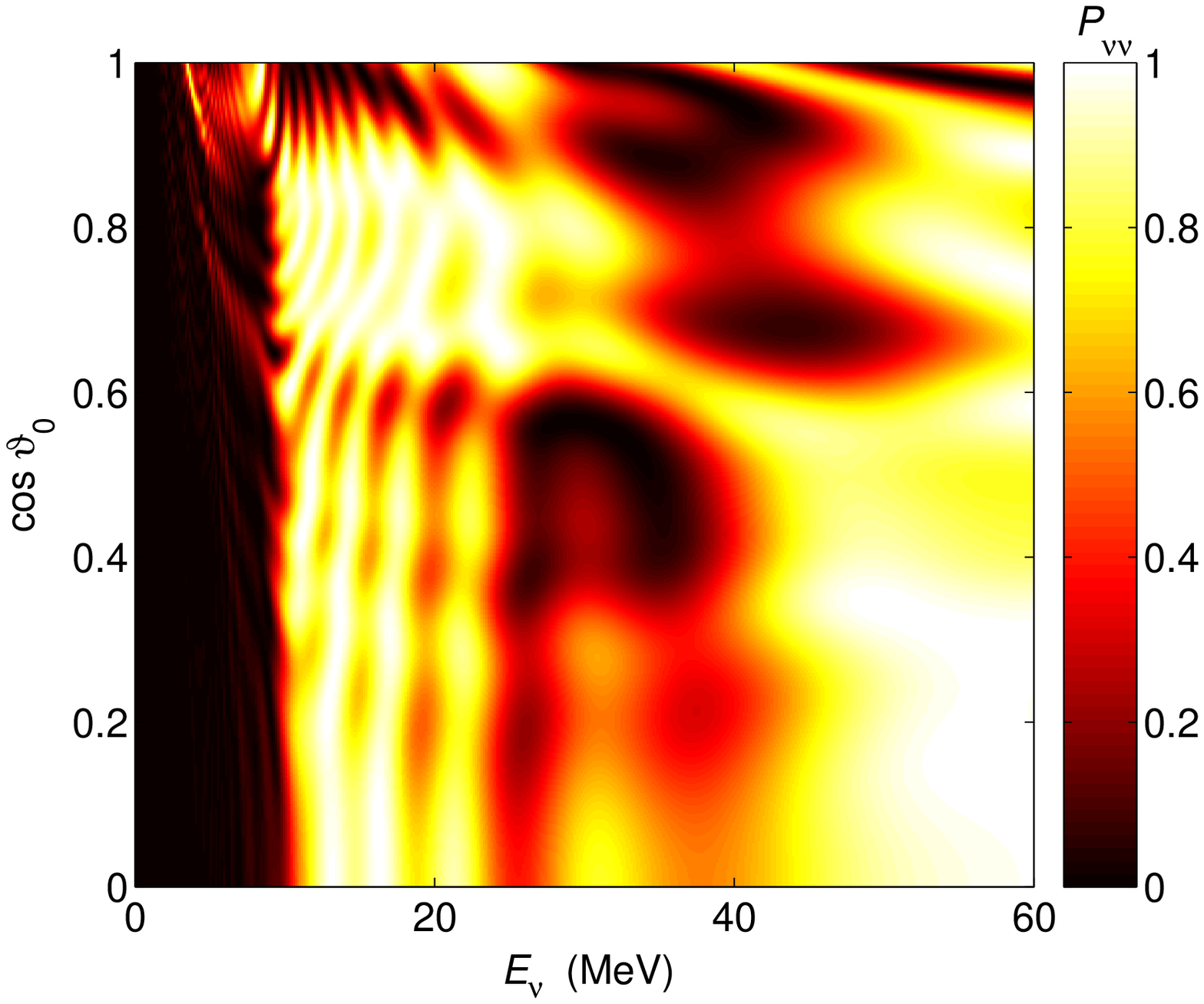} &
\includegraphics*[scale=\myfigscl, keepaspectratio]{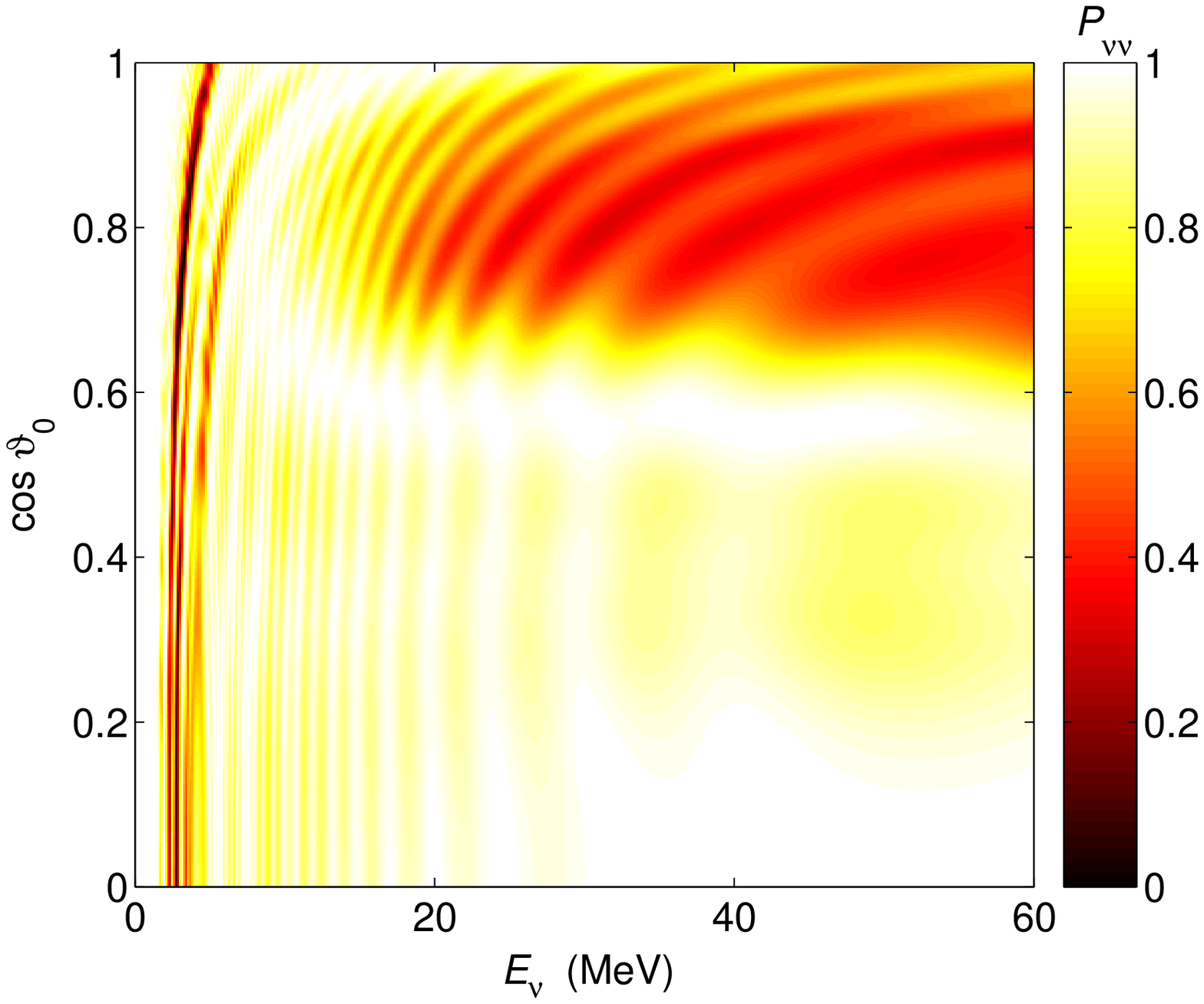} \\
\includegraphics*[scale=\myfigscl, keepaspectratio]{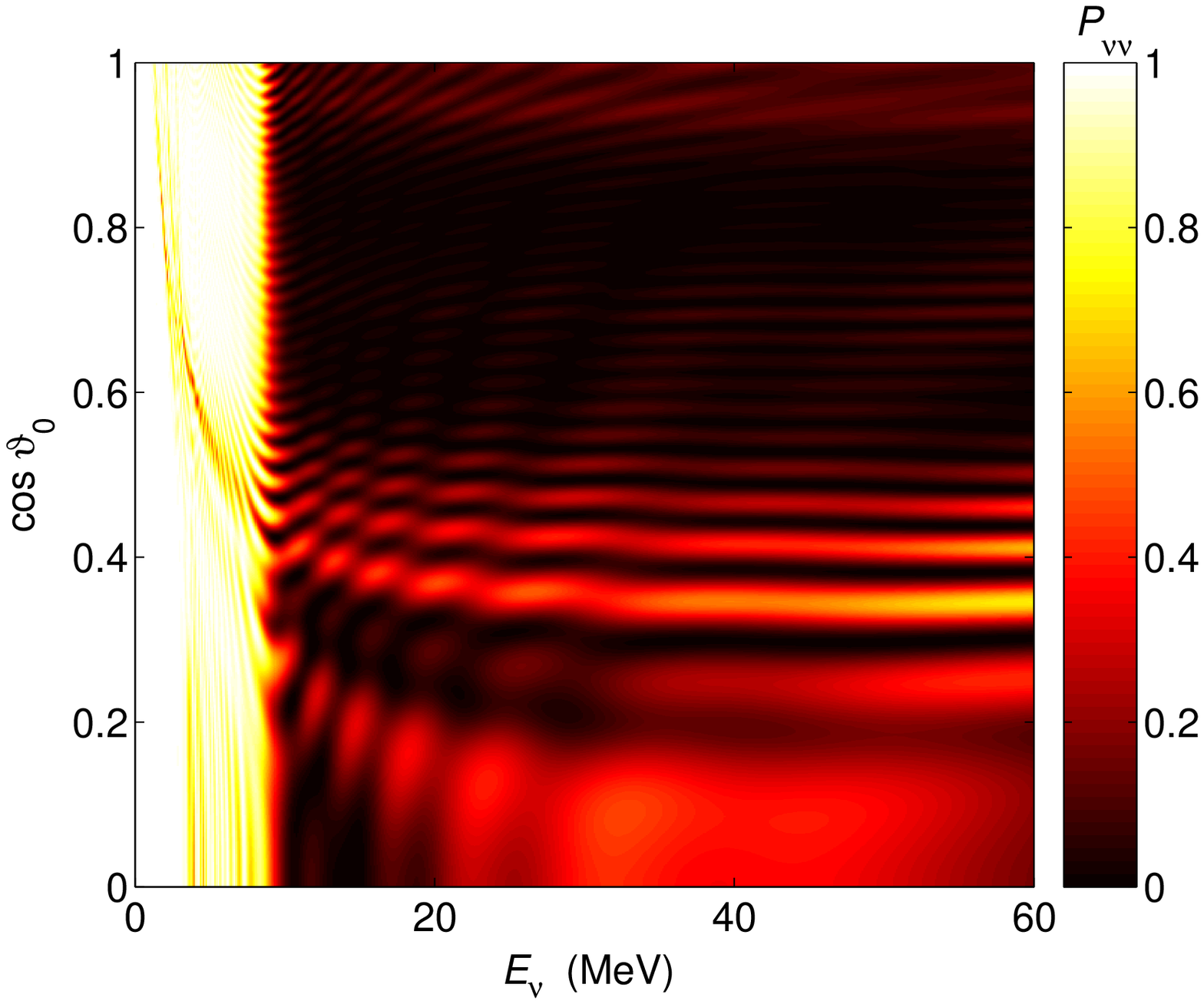} &
\includegraphics*[scale=\myfigscl, keepaspectratio]{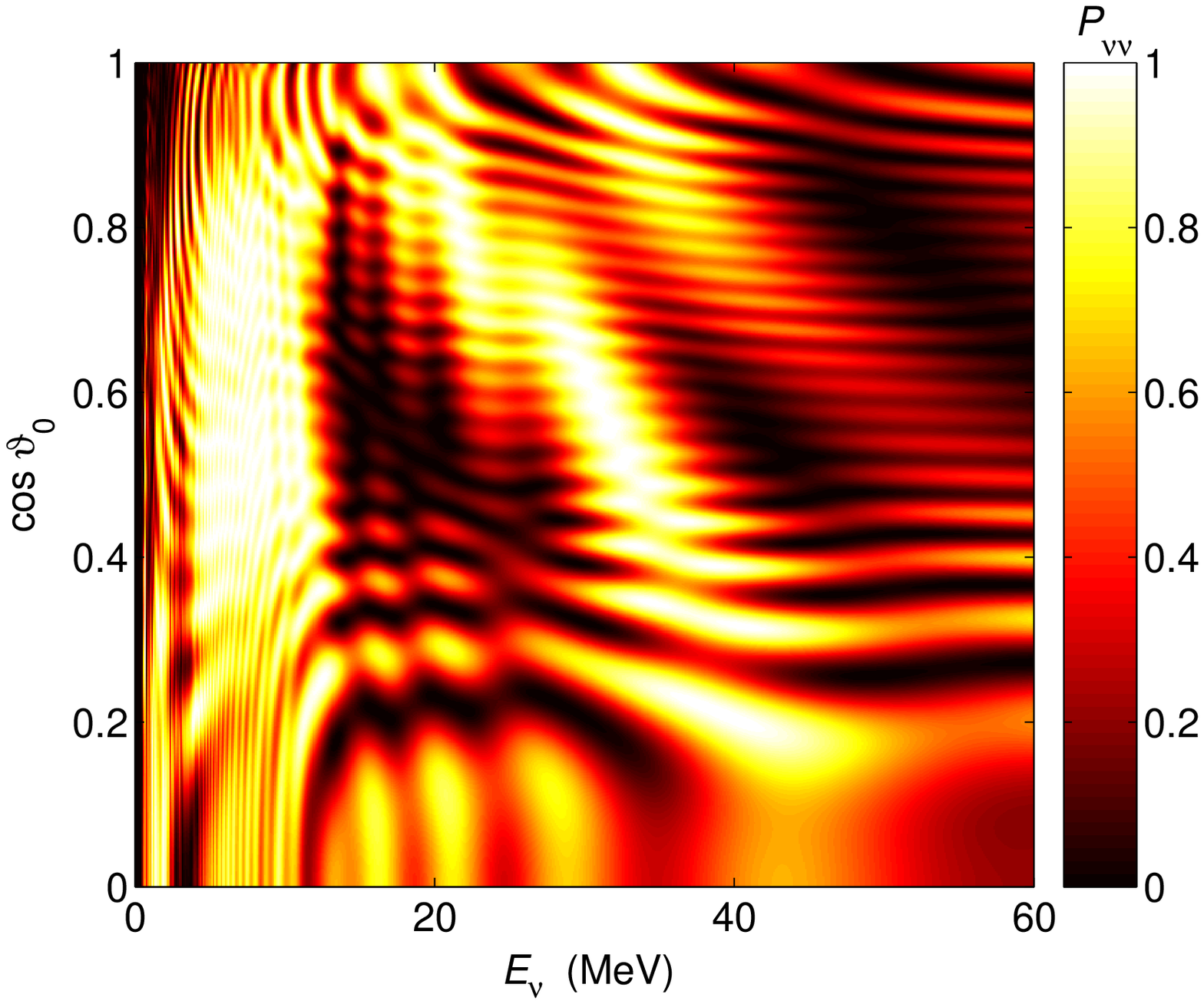}
\end{array}$
\end{center}
\caption{\label{fig:P-E-c}(Color online) 
Plots of survival probabilities $P_{\nu\nu}$ for neutrinos (left panels) and
antineutrinos (right panels) as functions of both neutrino energy $E_\nu$
and emission angle $\vartheta_0$ at radius $r=225$ km.
The upper panels employ a normal neutrino mass hierarchy,
and the lower panels employ an inverted neutrino mass hierarchy.}
\end{figure*}

Though neutrinos and antineutrinos are emitted from
the neutrino sphere (radius $r=11$ km) in a thermal, incoherent manner,
our simulations show that large scale coherent and collective
flavor transformation develops with increasing radius $r$. This behavior
is driven in part by progressive forward scattering-induced 
coupling of flavor evolution on intersecting neutrino
and antineutrino world lines. Very far from the neutron star surface, where
neutrino densities are low, $B$ and $B_{e\tau}$ are negligible,
and ordinary Mikheyev-Smirnov-Wolfenstein (MSW) 
\cite{Wolfenstein:1977ue,Wolfenstein:1979ni,Mikheyev:1985aa}
flavor evolution or vacuum oscillations obtain.
However, at smaller $r$, $B$ and $B_{e \tau}$ can dominate 
and flavor evolution can be very different from MSW.
For the normal mass hierarchy,  our simulations show
that the more tangentially-propagating neutrinos and antineutrinos, 
which experience the largest
$B$ and $B_{e \tau}$ potentials because of
the intersection angle
dependence in the weak current \cite{Fuller:1987aa,Qian:1994wh}, 
are the first to experience significant 
flavor transformation
for a broad range of energies. This sets in, {\it e.g.}, 
at $r\approx 80\,{\rm km}$
if the entropy-per-baryon in the hot bubble is $S = 140$
in units of Boltzmann's constant $k_{\rm B}$. 
This simultaneous conversion of $\nu_\alpha$ and $\bar\nu_\alpha$ 
quickly spreads to all neutrino/antineutrino trajectories, 
leading to coherent, collective flavor oscillations of the
entire neutrino/antineutrino field. 
For the inverted mass hierarchy, the opposite is true: radially-propagating
neutrinos and antineutrinos transform first.

These features of flavor development
can be seen in Fig.~\ref{fig:P-r}. 
The survival probability at location $t$ along a given neutrino's world line
is, {\it e.g.}, for a neutrino which is initially electron flavor,
$P_{\nu_e\nu_e}\left(t,{\vartheta}_0, E_\nu\right) =
{\vert a_{e e}\left( t\right)\vert}^2$.
In Fig.~\ref{fig:P-r} we show
the energy-spectrum-averaged survival 
probabilities $\langle P_{\nu\nu}\rangle$ for $\nu_e$ and
$\bar\nu_e$ as functions of $r$
for both the normal and the
inverted neutrino mass hierarchy cases.
Here the energy averages are over the initial
energy spectra for each flavor.
It is clear that flavor evolution along different trajectories
can be different, yet it is also evident
that neutrinos and antineutrinos can undergo simultaneous, 
significant medium-enhanced flavor conversion. Our simulations
show that this conversion can take place over broad
ranges of neutrino and antineutrino energy.
We have also performed simulations using the single-angle approximation
widely adopted in the literature. These give results
{\it qualitatively} similar to our multi-angle calculations, as shown in
Fig.~\ref{fig:P-r}.
The collective neutrino flavor transformation
observed in our simulations is not the ``synchronized'' mode
described in Ref.~\cite{Pastor:2002we}. In the normal mass hierarchy case,
neutrinos or antineutrinos in the synchronized mode
undergo one-time transformation in the same way as does a neutrino with
energy $p_{\mathrm{sync}}$ \cite{Pastor:2002we}.
There would be little synchronized
flavor transformation in the inverted neutrino mass hierarchy.

The collective neutrino flavor transformation
evident in Fig.~\ref{fig:P-r} 
is likely of the ``bi-polar'' type as
described in Ref.~\cite{Duan:2005cp}. In this mode, neutrinos and antineutrinos 
experience in-phase, collective, semi-periodic flavor oscillations,
even for the inverted mass hierarchy. This behavior
was first observed in numerical simulations of neutrino
flavor transformation in the early universe 
\cite{Kostelecky:1993dm,Kostelecky:1996bs}. 
It has been argued \cite{Duan:2005cp}
that neutrinos and antineutrinos could undergo collective
flavor transformation in the bi-polar mode for the
inverted mass hierarchy case even in the presence of a dominant
matter potential. This conjecture seems to be supported by our simulations.

Using the single-angle approximation, we have studied the relation
between $\rX$, the radius where large-scale flavor transformation
first occurs, and the entropy per baryon $S$ and neutrino luminosity $L_\nu$.
With larger $S$ the matter density profile is more condensed 
toward the neutrino sphere. 
We find that, for the normal neutrino mass hierarchy case,
$\rX$ decreases substantially if $S$ is increased
from $140$ to $250$. However, the value of $\rX$
decreases only slightly with the same change in $S$
for the inverted neutrino mass hierarchy case.
The non-linear effect of the neutrino-neutrino
forward scattering potentials is enhanced with higher 
$L_\nu$. For the normal neutrino mass hierarchy case,
$\rX$ decreases monotonically with increasing $L_\nu$, and
approaches the radius where neutrinos and antineutrinos
start collective flavor transformation in the
synchronized mode. For the inverted neutrino mass hierarchy,
$\rX$ increases monotonically with increasing $L_\nu$.


In Fig.~\ref{fig:P-E-c} we show survival probabilities $P_{\nu\nu}$ 
as functions of both neutrino energy $E_\nu$
and emission angle $\vartheta_0$ at radius $r=225$ km. 
Large-scale quantum interference
stemming from coupling of
flavor histories is evident. The horizontal ``fringes''
in Fig.~\ref{fig:P-E-c} are generated during 
neutrino/antineutrino background-dominated collective 
flavor transformation.
In this regime $E_\nu$ is unimportant, but $\vartheta_0$ is 
crucial. This is because the neutrino-neutrino forward 
scattering potentials are energy-blind but have a strong angle dependence.
The vertical \lq\lq fringes\rq\rq\ are generated at larger 
$r$ where the neutrino background is weaker and, therefore, 
(non-adiabatic) MSW transformation dominates flavor evolution. 
In this case $\vartheta_0$ is unimportant and $E_\nu$ is crucial. 
This is because at large $r$ the neutrino beams are almost parallel 
to each other, and the MSW effect is energy dependent.
Our simulations show sharp, vertical transition regions at
$E_\nu\simeq10$ MeV for neutrinos in both the normal and inverted
neutrino mass hierarchies. This feature is likely a result
of break-down of collective transformation \cite{Duan:2006an}.

Although the establishment and breakdown of collectivity
in  $\nu$ and $\bar\nu$ flavor transformation remains
an open issue for the general supernova environment, our particular
simulations are based on reasonable assumptions and our results
are robust with these assumptions. These simulations suggest
that, with $\delta m_\mathrm{atm}^2$,
large-scale collective flavor transformation in the bi-polar
mode can occur deep in the supernova envelope.
In broad brush, across 
{\it all} $\vartheta_0$ mostly lower energy $\nu_e$ and few 
$\bar\nu_e$ are transformed in the normal mass hierarchy. 
The opposite is true in the inverted mass hierarchy. 
However, as is evident in  Fig.~\ref{fig:P-E-c} at $r=225\,{\rm km}$, 
in either mass hierarchy, survival probability can 
show significant $\vartheta_0$ dependence. The $\nu_e$ and $\bar\nu_e$ 
energy spectra and angular distributions could be quite different from those
in zero-neutrino-mass supernova models.
This could affect the expected neutrino signal and conceivably
affect neutron-proton interconversion rates and the
prospects for $r$-process nucleosynthesis \cite{Qian:1993dg}.


\begin{acknowledgments}
This work was supported in part by a UC/LANL CARE grant,
NSF grant PHY-04-00359,
the Terascale Supernova Initiative (TSI) collaboration's 
DOE SciDAC grant at UCSD, 
DOE grant DE-FG02-87ER40328 at UMN,
and by the LDRD Program
and Open Supercomputing at LANL.
We would like to thank A.~B.~Balantekin, S.~Bruenn, 
C.~Y.~Cardall, J.~Hayes,
W.~Landry, O.~E.~B.~Messer, A.~Mezzacappa, M.~Patel, 
G.~Raffelt and H.~Y\"{u}ksel
for valuable conversations.
\end{acknowledgments}

\bibliography{ref}

\begin{thebibliography}{22}
\expandafter\ifx\csname natexlab\endcsname\relax\def\natexlab#1{#1}\fi
\expandafter\ifx\csname bibnamefont\endcsname\relax
  \def\bibnamefont#1{#1}\fi
\expandafter\ifx\csname bibfnamefont\endcsname\relax
  \def\bibfnamefont#1{#1}\fi
\expandafter\ifx\csname citenamefont\endcsname\relax
  \def\citenamefont#1{#1}\fi
\expandafter\ifx\csname url\endcsname\relax
  \def\url#1{\texttt{#1}}\fi
\expandafter\ifx\csname urlprefix\endcsname\relax\def\urlprefix{URL }\fi
\providecommand{\bibinfo}[2]{#2}
\providecommand{\eprint}[2][]{\url{#2}}

\bibitem[{\citenamefont{Duan et~al.}(2006)\citenamefont{Duan, Fuller, Carlson,
  and Qian}}]{Duan:2006an}
\bibinfo{author}{\bibfnamefont{H.}~\bibnamefont{Duan}},
  \bibinfo{author}{\bibfnamefont{G.~M.} \bibnamefont{Fuller}},
  \bibinfo{author}{\bibfnamefont{J.}~\bibnamefont{Carlson}}, \bibnamefont{and}
  \bibinfo{author}{\bibfnamefont{Y.-Z.} \bibnamefont{Qian}}
  (\bibinfo{year}{2006}), \eprint{astro-ph/0606616}.

\bibitem[{\citenamefont{Qian and Fuller}(1995{\natexlab{a}})}]{Qian:1994wh}
\bibinfo{author}{\bibfnamefont{Y.~Z.} \bibnamefont{Qian}} \bibnamefont{and}
  \bibinfo{author}{\bibfnamefont{G.~M.} \bibnamefont{Fuller}},
  \bibinfo{journal}{Phys. Rev.} \textbf{\bibinfo{volume}{D51}},
  \bibinfo{pages}{1479} (\bibinfo{year}{1995}{\natexlab{a}}),
  \eprint{astro-ph/9406073}.

\bibitem[{\citenamefont{Fuller et~al.}(1987)\citenamefont{Fuller, Mayle,
  Wilson, and Schramm}}]{Fuller:1987aa}
\bibinfo{author}{\bibfnamefont{G.~M.} \bibnamefont{Fuller}},
  \bibinfo{author}{\bibfnamefont{R.~W.} \bibnamefont{Mayle}},
  \bibinfo{author}{\bibfnamefont{J.~R.} \bibnamefont{Wilson}},
  \bibnamefont{and} \bibinfo{author}{\bibfnamefont{D.~N.}
  \bibnamefont{Schramm}}, \bibinfo{journal}{Astrophys. J.}
  \textbf{\bibinfo{volume}{322}}, \bibinfo{pages}{795} (\bibinfo{year}{1987}).

\bibitem[{\citenamefont{Fuller et~al.}(1992)\citenamefont{Fuller, Mayle, Meyer,
  and Wilson}}]{Fuller:1992aa}
\bibinfo{author}{\bibfnamefont{G.~M.} \bibnamefont{Fuller}},
  \bibinfo{author}{\bibfnamefont{R.~W.} \bibnamefont{Mayle}},
  \bibinfo{author}{\bibfnamefont{B.~S.} \bibnamefont{Meyer}}, \bibnamefont{and}
  \bibinfo{author}{\bibfnamefont{J.~R.} \bibnamefont{Wilson}},
  \bibinfo{journal}{Astrophys. J.} \textbf{\bibinfo{volume}{389}},
  \bibinfo{pages}{517} (\bibinfo{year}{1992}).

\bibitem[{\citenamefont{Mezzacappa and Bruenn}(1999)}]{Mezzacappa:1999co}
\bibinfo{author}{\bibfnamefont{A.}~\bibnamefont{Mezzacappa}} \bibnamefont{and}
  \bibinfo{author}{\bibfnamefont{S.}~\bibnamefont{Bruenn}}, in
  \emph{\bibinfo{booktitle}{Proceeding of the Second International Workshop on
  the Identification of Dark Matter}}, edited by
  \bibinfo{editor}{\bibfnamefont{N.~J.~C.} \bibnamefont{Spooner}}
  \bibnamefont{and}
  \bibinfo{editor}{\bibfnamefont{V.}~\bibnamefont{Kudryavtsev}}
  (\bibinfo{publisher}{World Scientific}, \bibinfo{address}{Singopore},
  \bibinfo{year}{1999}).

\bibitem[{\citenamefont{Qian et~al.}(1993)\citenamefont{Qian, Fuller, Mathews,
  Mayle, Wilson, and Woosley}}]{Qian:1993dg}
\bibinfo{author}{\bibfnamefont{Y.-Z.} \bibnamefont{Qian}},
  \bibinfo{author}{\bibfnamefont{G.~M.} \bibnamefont{Fuller}},
  \bibinfo{author}{\bibfnamefont{G.~J.} \bibnamefont{Mathews}},
  \bibinfo{author}{\bibfnamefont{R.~W.} \bibnamefont{Mayle}},
  \bibinfo{author}{\bibfnamefont{J.~R.} \bibnamefont{Wilson}},
  \bibnamefont{and} \bibinfo{author}{\bibfnamefont{S.~E.}
  \bibnamefont{Woosley}}, \bibinfo{journal}{Phys. Rev. Lett.}
  \textbf{\bibinfo{volume}{71}}, \bibinfo{pages}{1965} (\bibinfo{year}{1993}).

\bibitem[{\citenamefont{Pastor and Raffelt}(2002)}]{Pastor:2002we}
\bibinfo{author}{\bibfnamefont{S.}~\bibnamefont{Pastor}} \bibnamefont{and}
  \bibinfo{author}{\bibfnamefont{G.}~\bibnamefont{Raffelt}},
  \bibinfo{journal}{Phys. Rev. Lett.} \textbf{\bibinfo{volume}{89}},
  \bibinfo{pages}{191101} (\bibinfo{year}{2002}), \eprint{astro-ph/0207281}.

\bibitem[{\citenamefont{Qian and Fuller}(1995{\natexlab{b}})}]{Qian:1995ua}
\bibinfo{author}{\bibfnamefont{Y.-Z.} \bibnamefont{Qian}} \bibnamefont{and}
  \bibinfo{author}{\bibfnamefont{G.~M.} \bibnamefont{Fuller}},
  \bibinfo{journal}{Phys. Rev.} \textbf{\bibinfo{volume}{D52}},
  \bibinfo{pages}{656} (\bibinfo{year}{1995}{\natexlab{b}}),
  \eprint{astro-ph/9502080}.

\bibitem[{\citenamefont{Balantekin and Y\"{u}ksel}(2005)}]{Balantekin:2004ug}
\bibinfo{author}{\bibfnamefont{A.~B.} \bibnamefont{Balantekin}}
  \bibnamefont{and}
  \bibinfo{author}{\bibfnamefont{H.}~\bibnamefont{Y\"{u}ksel}},
  \bibinfo{journal}{New J. Phys.} \textbf{\bibinfo{volume}{7}},
  \bibinfo{pages}{51} (\bibinfo{year}{2005}), \eprint{astro-ph/0411159}.

\bibitem[{\citenamefont{Fuller and Qian}(2006)}]{Fuller:2005ae}
\bibinfo{author}{\bibfnamefont{G.~M.} \bibnamefont{Fuller}} \bibnamefont{and}
  \bibinfo{author}{\bibfnamefont{Y.-Z.} \bibnamefont{Qian}},
  \bibinfo{journal}{Phys. Rev.} \textbf{\bibinfo{volume}{D73}},
  \bibinfo{pages}{023004} (\bibinfo{year}{2006}), \eprint{astro-ph/0505240}.

\bibitem[{\citenamefont{Sigl and Raffelt}(1993)}]{Sigl:1992fn}
\bibinfo{author}{\bibfnamefont{G.}~\bibnamefont{Sigl}} \bibnamefont{and}
  \bibinfo{author}{\bibfnamefont{G.}~\bibnamefont{Raffelt}},
  \bibinfo{journal}{Nucl. Phys.} \textbf{\bibinfo{volume}{B406}},
  \bibinfo{pages}{423} (\bibinfo{year}{1993}).

\bibitem[{\citenamefont{Strack and Burrows}(2005)}]{Strack:2005ux}
\bibinfo{author}{\bibfnamefont{P.}~\bibnamefont{Strack}} \bibnamefont{and}
  \bibinfo{author}{\bibfnamefont{A.}~\bibnamefont{Burrows}},
  \bibinfo{journal}{Phys. Rev.} \textbf{\bibinfo{volume}{D71}},
  \bibinfo{pages}{093004} (\bibinfo{year}{2005}), \eprint{hep-ph/0504035}.

\bibitem[{\citenamefont{Friedland and Lunardini}(2003)}]{Friedland:2003eh}
\bibinfo{author}{\bibfnamefont{A.}~\bibnamefont{Friedland}} \bibnamefont{and}
  \bibinfo{author}{\bibfnamefont{C.}~\bibnamefont{Lunardini}},
  \bibinfo{journal}{JHEP} \textbf{\bibinfo{volume}{10}}, \bibinfo{pages}{043}
  (\bibinfo{year}{2003}), \eprint{hep-ph/0307140}.

\bibitem[{\citenamefont{Fogli et~al.}(2006)\citenamefont{Fogli, Lisi, Marrone,
  and Palazzo}}]{Fogli:2005cq}
\bibinfo{author}{\bibfnamefont{G.~L.} \bibnamefont{Fogli}},
  \bibinfo{author}{\bibfnamefont{E.}~\bibnamefont{Lisi}},
  \bibinfo{author}{\bibfnamefont{A.}~\bibnamefont{Marrone}}, \bibnamefont{and}
  \bibinfo{author}{\bibfnamefont{A.}~\bibnamefont{Palazzo}},
  \bibinfo{journal}{Prog. Part. Nucl. Phys.} \textbf{\bibinfo{volume}{57}},
  \bibinfo{pages}{742} (\bibinfo{year}{2006}), \eprint{hep-ph/0506083}.

\bibitem[{\citenamefont{Pantaleone}(1992)}]{Pantaleone:1992xh}
\bibinfo{author}{\bibfnamefont{J.~T.} \bibnamefont{Pantaleone}},
  \bibinfo{journal}{Phys. Rev.} \textbf{\bibinfo{volume}{D46}},
  \bibinfo{pages}{510} (\bibinfo{year}{1992}).

\bibitem[{\citenamefont{Balantekin and Fuller}(1999)}]{Balantekin:1999dx}
\bibinfo{author}{\bibfnamefont{A.~B.} \bibnamefont{Balantekin}}
  \bibnamefont{and} \bibinfo{author}{\bibfnamefont{G.~M.}
  \bibnamefont{Fuller}}, \bibinfo{journal}{Phys. Lett.}
  \textbf{\bibinfo{volume}{B471}}, \bibinfo{pages}{195} (\bibinfo{year}{1999}),
  \eprint{hep-ph/9908465}.

\bibitem[{\citenamefont{Wolfenstein}(1978)}]{Wolfenstein:1977ue}
\bibinfo{author}{\bibfnamefont{L.}~\bibnamefont{Wolfenstein}},
  \bibinfo{journal}{Phys. Rev.} \textbf{\bibinfo{volume}{D17}},
  \bibinfo{pages}{2369} (\bibinfo{year}{1978}).

\bibitem[{\citenamefont{Wolfenstein}(1979)}]{Wolfenstein:1979ni}
\bibinfo{author}{\bibfnamefont{L.}~\bibnamefont{Wolfenstein}},
  \bibinfo{journal}{Phys. Rev.} \textbf{\bibinfo{volume}{D20}},
  \bibinfo{pages}{2634} (\bibinfo{year}{1979}).

\bibitem[{\citenamefont{Mikheyev and Smirnov}(1985)}]{Mikheyev:1985aa}
\bibinfo{author}{\bibfnamefont{S.~P.} \bibnamefont{Mikheyev}} \bibnamefont{and}
  \bibinfo{author}{\bibfnamefont{A.~Y.} \bibnamefont{Smirnov}},
  \bibinfo{journal}{Yad. Fiz.} \textbf{\bibinfo{volume}{42}},
  \bibinfo{pages}{1441} (\bibinfo{year}{1985}).

\bibitem[{\citenamefont{Duan et~al.}(2005)\citenamefont{Duan, Fuller, and
  Qian}}]{Duan:2005cp}
\bibinfo{author}{\bibfnamefont{H.}~\bibnamefont{Duan}},
  \bibinfo{author}{\bibfnamefont{G.~M.} \bibnamefont{Fuller}},
  \bibnamefont{and} \bibinfo{author}{\bibfnamefont{Y.-Z.} \bibnamefont{Qian}}
  (\bibinfo{year}{2005}), \eprint{astro-ph/0511275}.

\bibitem[{\citenamefont{Kostelecky and Samuel}(1993)}]{Kostelecky:1993dm}
\bibinfo{author}{\bibfnamefont{V.~A.} \bibnamefont{Kostelecky}}
  \bibnamefont{and} \bibinfo{author}{\bibfnamefont{S.}~\bibnamefont{Samuel}},
  \bibinfo{journal}{Phys. Lett.} \textbf{\bibinfo{volume}{B318}},
  \bibinfo{pages}{127} (\bibinfo{year}{1993}).

\bibitem[{\citenamefont{Kostelecky and Samuel}(1996)}]{Kostelecky:1996bs}
\bibinfo{author}{\bibfnamefont{V.~A.} \bibnamefont{Kostelecky}}
  \bibnamefont{and} \bibinfo{author}{\bibfnamefont{S.}~\bibnamefont{Samuel}},
  \bibinfo{journal}{Phys. Lett.} \textbf{\bibinfo{volume}{B385}},
  \bibinfo{pages}{159} (\bibinfo{year}{1996}), \eprint{hep-ph/9610399}.

\end{thebibliography}

\end{document}